\documentclass[runningheads]{llncs}

\usepackage{amssymb,graphicx,subfigure,url}
\newcommand{\keywords}[1]{\par\addvspace\baselineskip
\noindent\keywordname\enspace\ignorespaces#1}

\begin{document}

\mainmatter  

\title{Real-time fMRI-based Brain Computer Interface: A Review}
\titlerunning{Real-time fMRI-based Brain Computer Interface: A Review}

\author{Yang Wang and Dongrui Wu}
\authorrunning{Yang Wang and Dongrui Wu}
\institute{School of Automation\\ Huazhong University of Science and Technology, Wuhan, Hubei, China\\
\email{wangyang\_sky@hust.edu.cn}, \email{drwu@hust.edu.cn}}
\maketitle

\begin{abstract}
In recent years, the rapid development of neuroimaging technology has been providing many powerful tools for cognitive neuroscience research. Among them, the functional magnetic resonance imaging (fMRI), which has high spatial resolution, acceptable temporal resolution, simple calibration, and short preparation time, has been widely used in brain research. Compared with the electroencephalogram (EEG), real-time fMRI-based brain computer interface (rtfMRI-BCI) not only can perform decoding analysis across the whole brain to control external devices, but also allows a subject to voluntarily self-regulate specific brain regions. This paper reviews the basic architecture of rtfMRI-BCI, the emerging machine learning based data analysis approaches (also known as multi-voxel pattern analysis), and the applications and recent advances of rtfMRI-BCI.

\end{abstract}

\section{Introduction}

A brain computer interface (BCI) uses neurophysiological signals from the brain, e.g., electrocorticography (ECoG), electroencephalogram (EEG), and functional magnetic resonance imaging (fMRI), to control external devices or computers \cite{1}. Among these signals, fMRI non-invasively measures the task-induced blood-oxygen-level-dependent (BOLD) changes related to brain neuronal activities. Unlike EEG, fMRI has excellent spatial resolution and whole brain coverage, so it can accurately locate activation areas in the brain.

This paper reviews the basic architecture of real-time fMRI-based BCI (rtfMRI-BCI), an emerging machine learning based data analysis approach (also known as multi-voxel pattern analysis), and the applications and recent advances of rtfMRI-BCI.

\section{The Architecture of rtfMRI-BCI}

Different from conventional fMRI, in which image analysis can only be performed after all scans are finished, rtfMRI-based BCI allows the simultaneous acquisition, analysis and visualization of whole brain images. A typical closed-loop rtfMRI-BCI system consists of four components: image acquisition, image preprocessing, image analysis, and feedback.

\begin{enumerate}
\item \emph{Image acquisition}: According to some pre-defined scanning parameters, a MRI scanner uses an echo planar imaging sequence to stimulate brain MRI echo signals and then records them. An image reconstruction workstation then assembles these signals into three-dimensional images.

\item \emph{Image preprocessing}: fMRI images need to be preprocessed to improve their quality before further analyses can be performed. This usually involves the following steps:
\begin{enumerate}
\item \emph{Slice timing correction}: An fMRI image consists of multiple slices that are sampled sequentially at different time instances, so the same region from different slices are shifted in time relative to each other. Slice timing correction interpolates the slices so that they can be viewed as being sampled at exactly the same time \cite{SLADKY_Slice}, as shown in Fig.~\ref{fig:slice}.

\item \emph{Realignment}: Any head motion of the subject can contaminate the neighboring voxels. A common practice for motion correction is to treat the brain as a rigid body, and then calculate its translation and rotation relative to a reference image \cite{Friston_head}.

\item \emph{Coregistration}: fMRI images typically have low spatial resolution and do not include enough anatomical details, so they are usually registered to a high resolution structural MRI image of the same subject before presentation \cite{Wells_registration}.

\item \emph{Normalization}: Group analysis requires the voxels from the same brain location of different subjects are comparable. Normalization is used to register a subject's anatomical structure to a standardized stereotaxic space defined by a template, such as the Montreal Neurological Institute or Talairach brain \cite{Ashburner_normalization}.

\item \emph{Spatial smoothing}: This is usually performed by convolving the functional image with a Gaussian kernel. Smoothing can suppress random noise, and hence increase the signal-to-noise ratio. However, it also reduces the actual spatial resolution and blurs the details, so generally it is not used in machine learning based fMRI analysis.
\end{enumerate}

\begin{figure}\centering
\includegraphics[width=.55\linewidth,clip]{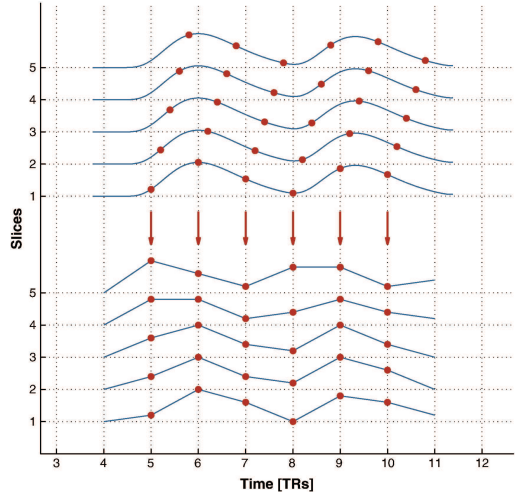}
\caption{Illustration of slice timing correction. Adopted from \cite{SLADKY_Slice}.} \label{fig:slice}
\end{figure}

\item \emph{Image analysis}: This step locates the real-time activation areas within the brain and then performs univariate or multivariate analysis. Typical tasks include statistical analysis of a specific region of interest (ROI) to determine its activation level, and online classification of brain states to find the subject's intention.

Univariate analysis measures brain activities from thousands of locations repeatedly, and then analyzes each location individually to understand how a particular perceptual or cognitive state is encoded \cite{2}. If the response at a certain location in the brain is different between two states, then the voxel strength at that location can be used to decode the state. Therefore, univariate analysis uses statistical analysis to identify the voxels that are significantly correlated to a specific task, and hence the regions that are significantly activated in the brain, which are called ROIs or functional areas.

While the majority of work in rtfMRI-BCI is done through conventional univariate analysis, there is a growing interest in machine learning based multivariate analysis, particularly, in the emerging field of brain state classification, i.e., decoding the brain state to determine the intention of the subject. This typically includes feature extraction, feature selection/dimensionality reduction, and classification.

\begin{enumerate}
\item \emph{Feature extraction}: The resting-state fMRI is commonly used to diagnose mental diseases. In addition to calculating regional attributes such as the amplitude of low-frequency fluctuations \cite{3} and regional homogeneity \cite{4}, functional connections between different regions can also be calculated, and the connection matrix can be used to compute its network properties \cite{5}. For the task-based fMRI, in addition to calculating the functional connections between different regions, the voxel intensities at different times can also be used as features in pattern analysis, and the resulting method is called multi-voxel pattern analysis (MVPA).

\item \emph{Feature selection/dimensionality reduction}: Feature selection selects the most useful features from a feature set and discards the rest, so it also results in dimensionality reduction. It is an important data preprocessing process that can alleviate the curse of dimensionality and simplify the subsequent learning tasks. Dimensionality reduction maps the original high-dimensional feature space to a low-dimensional subspace using a mathematical transformation. The new features are linear or nonlinear combinations of the original features, and are usually more informative \cite{6}.

\item \emph{Classification}: Simple linear classifiers, such as correlation-based classifier \cite{7,8}, neural networks without hidden layers \cite{9}, linear discriminant analysis \cite{10,11,12,13}, linear support vector machine (SVM) \cite{14,15,16}, and Gaussian naive Bayes classifiers \cite{14}, are frequently used in MVPA. They compute a weighted sum of the voxel intensities and pass it to a decision function to classify the brain state. Nonlinear classifiers, such as nonlinear SVM \cite{14,17} and multi-layer neural networks \cite{18}, have also been used in MVPA. Compared with linear classifiers, nonlinear ones can capture more complex mappings between features and the brain states. Though theoretically nonlinear classifiers can implement more complex mappings, there is no guarantee that they can significantly outperform linear classifiers in MVPA \cite{15}. This may be because nonlinear classifiers generally need a large amount of training data to achieve their best performance, which may not be easily available in neuroimaging. Additionally, by using a simple linear classifier one can visualize and explain which voxels are more important in decision making, but it is much more difficult to do so for a nonlinear classifier. As a result, the linear SVM classifier is frequently used in fMRI research.
\end{enumerate}

\item \emph{Feedback}: This step feeds the online analysis results back to the subject in real-time, so that the subject can voluntarily self-regulate his/her cognitive function or state. It also presents task-related stimuli to the subject.
\end{enumerate}

\section{Applications of rtfMRI-BCI}

The applications of rtfMRI in BCI can be roughly partitioned into two categories: 1) neurofeedback, in which a subject can voluntarily self-regulate his/her brain activity in a specific region through the feedback of the activation level there; and, 2) brain state decoding, which analyzes the subject's fMRI data to determine his/her intention, which can be then used to control an external device or computer.

\subsection{Neurofeedback}

\begin{figure}\centering
\subfigure[]{\includegraphics[width=.3\linewidth,clip]{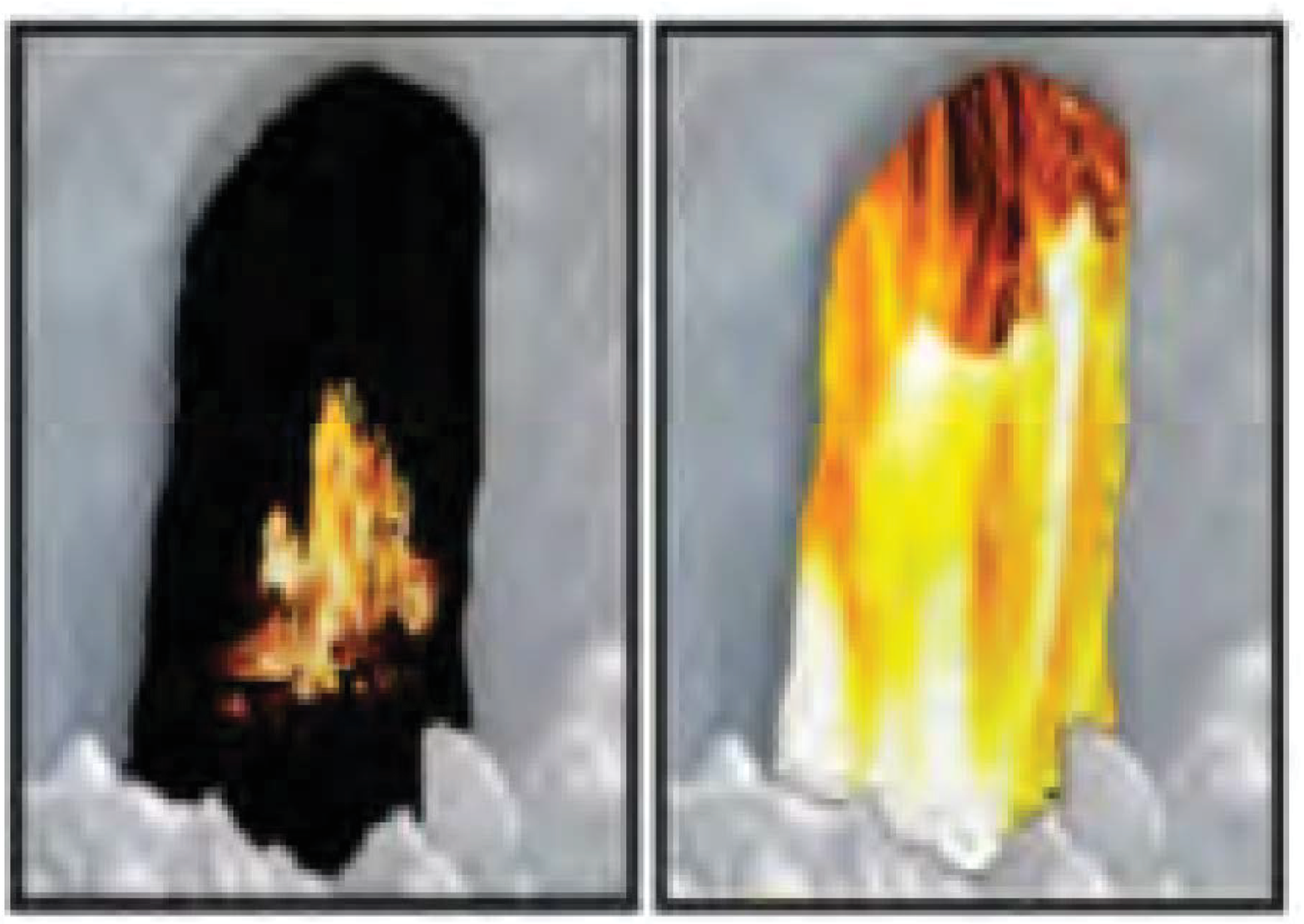}\label{fig:flame} }
\subfigure[]{\includegraphics[width=.315\linewidth,clip]{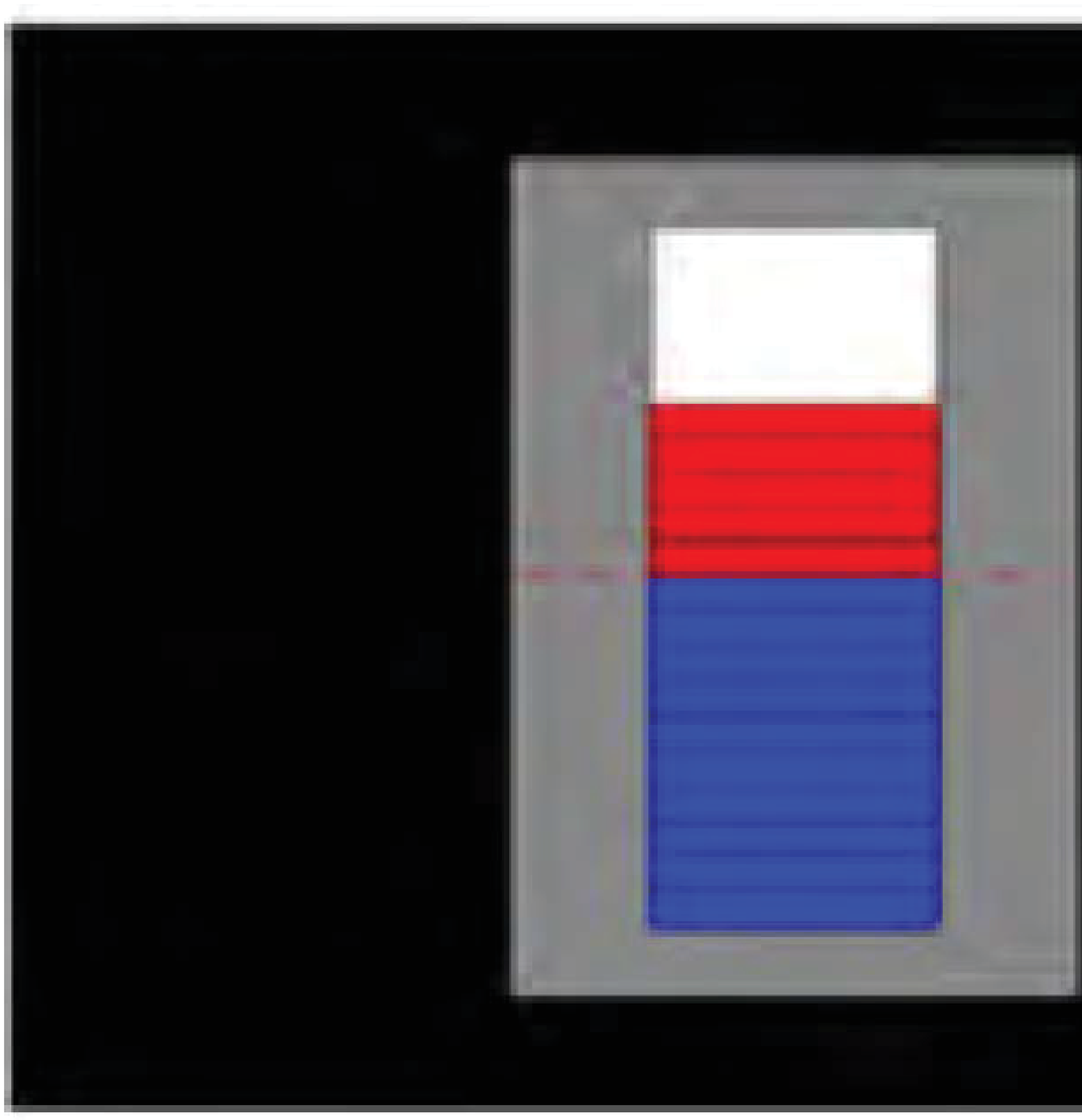}\label{fig:thermometer} }
\subfigure[]{\includegraphics[width=.29\linewidth,clip]{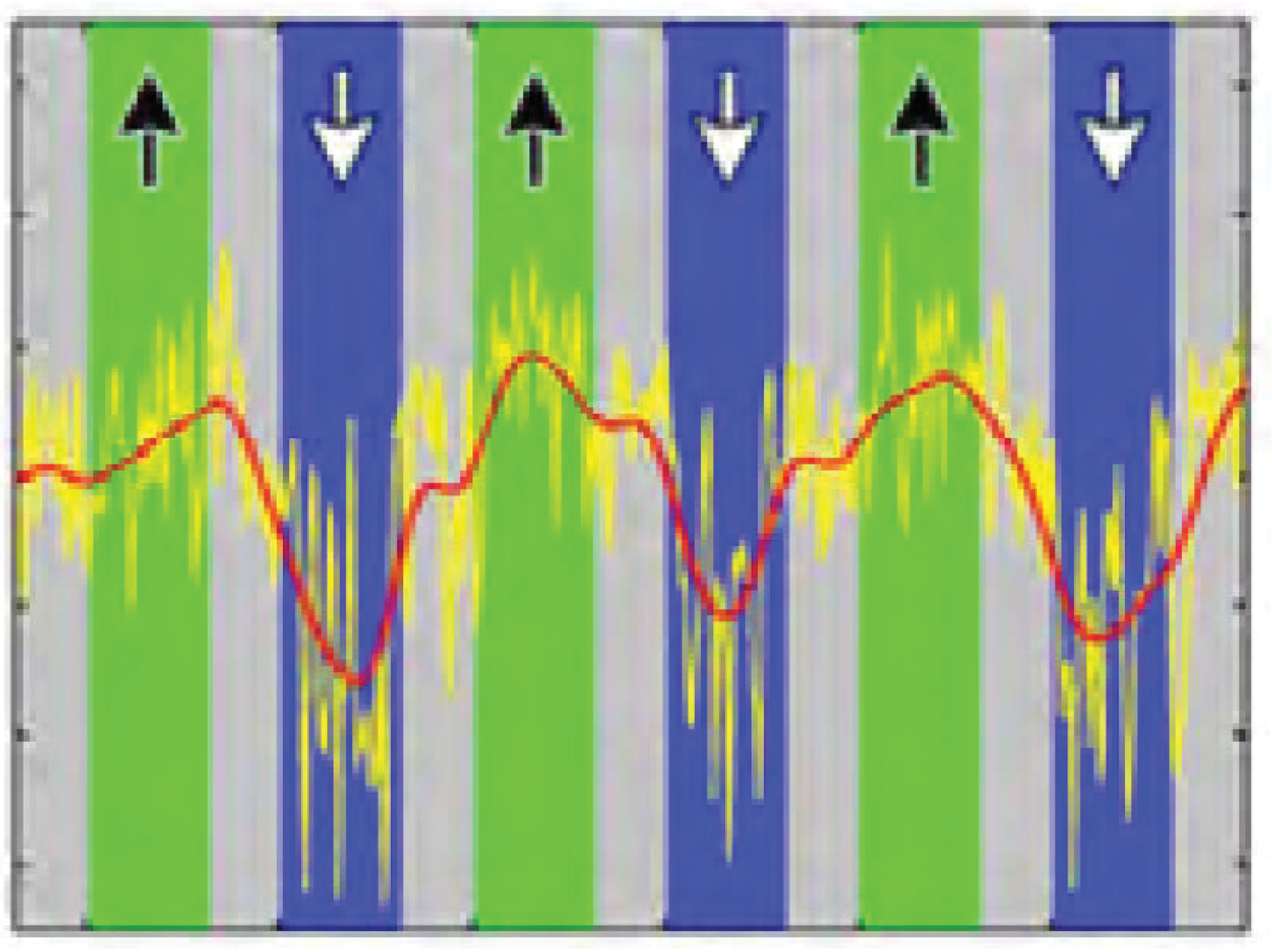}\label{fig:curve} }
\caption{Three different forms of visual feedback. (a) flame, adopted from \cite{19}; (b) thermometer, adopted from \cite{21}; (c) intensity curve, adopted from \cite{22}.}\label{fig:feedback}
\end{figure}

Because fMRI has high spatial resolution and can image the entire brain, rtfMRI-BCI can extract the activation levels of specific anatomical locations (ROIs) as feedback. Among the various feedback modalities (auditory, visual, verbal, olfactory, and tactile), visual feedback has been the most popular one. The form of visual feedback also changes with the purpose of the experiment. deCharms et al. \cite{19} introduced a flame-like feedback in a pain-related study, as shown in Fig.~\ref{fig:flame}, where the intensity of the flame increases with the intensity of the signal. Sitaram et al. \cite{21} described a thermometer feedback, where red and blue colors are used to indicate whether the signal is above or below a baseline, as shown in Fig.~\ref{fig:thermometer}. Weiskopf et al. \cite{22} used the differential feedback intensity curve as feedback, where an upward arrow indicates an activity enhancement, as shown in Fig.~\ref{fig:curve}.

The seminal rtfMRI-BCI work by deCarms et al. \cite{19} on chronic pain is worth special mentioning here. The purpose was to find out whether adjusting the activity on the rostral part of the anterior cingulate cortex (rACC) can affect the perception of pain. Their study showed that the pain introduced by noxious stimulus may be perceived differently if the subject intentionally induces an increase or inhibition in the BOLD level of rACC. Through  rtfMRI-based neurofeedback, subsequent experiments have been able to voluntarily adjust the level of activity in many other brain regions, including the anterior cingulate cortex \cite{20}, the insula \cite{21}, the motor area \cite{24}, the amygdala \cite{25}, the inferior frontal gyrus \cite{26}, and the parahippocampal place area \cite{27}. After enough training, a subject can even voluntarily adjust the corresponding brain region without neurofeedback, and this ability can last for some time after the training.

These research results suggest that rtfMRI-BCI provides a new approach in neuroscience for studying brain plasticity and functional reorganization through sustained training of specific brain regions \cite{28}. One potential application of neurofeedback is clinical rehabilitation, e.g., reducing the effects of abnormal brain activities, overcoming stroke-induced dyskinesia and Parkinson, relieving chronic pain, and treating depression and other neurological problems such as psychosis, social phobia and addiction \cite{30,31,32,33,34,35}.

\subsection{Brian State Decoding}

Another main application of rtfMRI-BCI is similar to ``brain reading", which classifies a subject's brain state to determine his/her intention. Its implementation can be divided into two categories: 1) pattern matching based on task-specific ROIs, and 2) machine learning based brain state classification.

Pattern matching was used by Yoo et al. \cite{36} in 2004 to perform BCI-based spatial navigation, in which a subject's brain signal was classified into four states so that they can control the computer to navigate through a maze. \cite{37,38,39,40,41} reported similar work. In all these studies the number of classifiable brain states did not exceed four.

In 2007, Sorger et al. \cite{42} used pattern matching to distinguish among 27 brain states, and implemented the world's first rtfMRI-BCI based spelling system. In this system, a subject can independently alter three aspects of the BOLD signal:
\begin{enumerate}
\item The location of the signal source, by performing three different mental tasks (motor imagery, mental calculation, and inner speech).
\item Delay of the mental task start time (0s, 10s, and 20s).
\item The duration of the mental task, which in turn determines the duration of the brain signal (10s, 20s, and 30s).
\end{enumerate}

\begin{figure}\centering
\includegraphics[height=4.05cm]{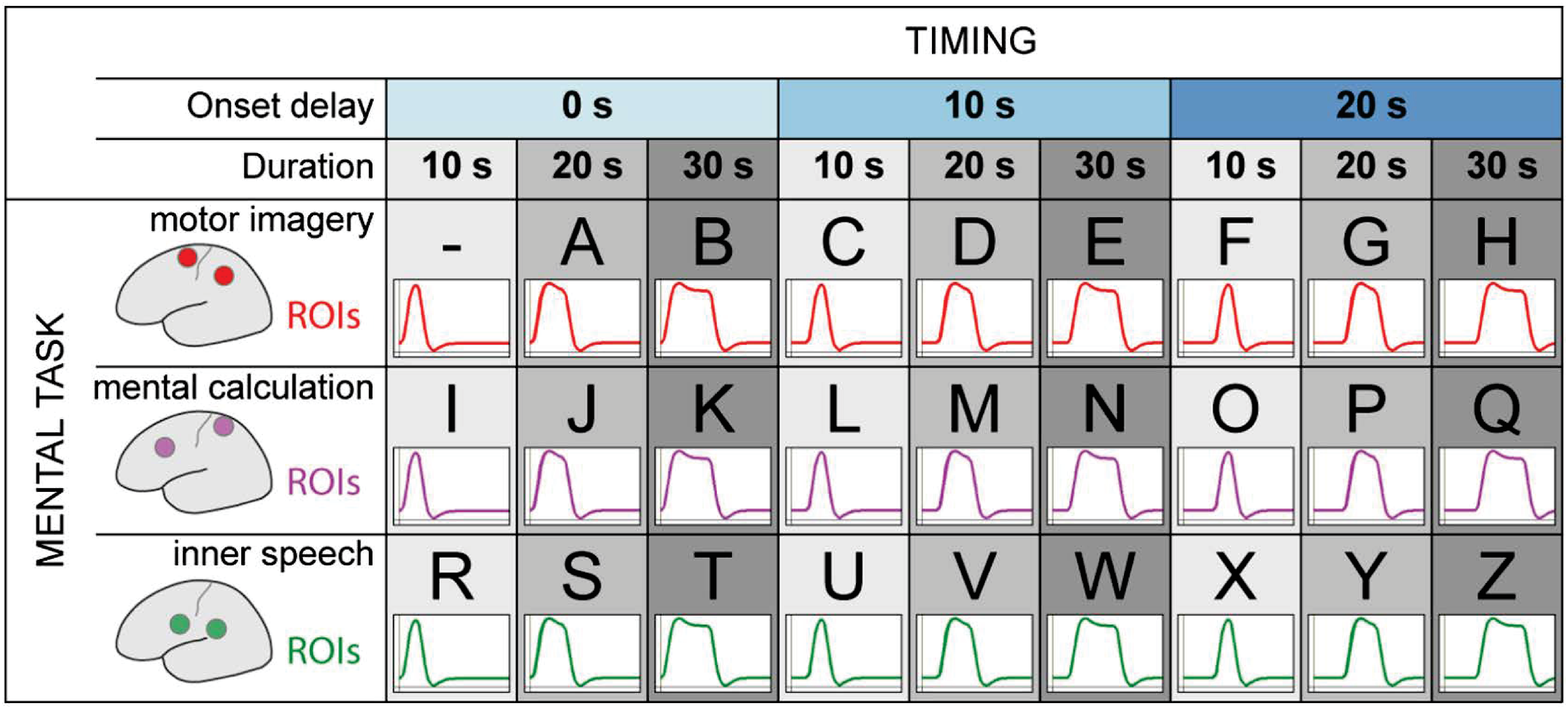}
\caption{Letter coding scheme. Adopted from \cite{42}.} \label{fig:coding}
\end{figure}

The combination of these aspects resulted in 27 unique brain responses, which can be assigned to 27 characters, as shown in Fig.~\ref{fig:coding}.

The spelling system required very little pre-training to help patients in locked-in syndrome to communicate in real time. Its main disadvantage is that the information transfer rate was very low (on average 50s per letter).

In summary, pattern matching based on task-specific ROIs needs very little pre-training and preparation to implement a BCI system, but generally has low transfer efficiency. Machine learning based brain state classification, also known as MVPA, is expected to improve it. Its main advantages include: 1) it does not require \emph{a priori} assumptions about the functional positioning and individual performance strategies, and 2) it can significantly improve the sensitivity of human neuroimaging analysis by considering the full spatial pattern of brain activities that are measured at many locations.

The application of MVPA to offline fMRI data analysis originated from Haxby et al.'s work \cite{7} in 2001. Since then, cognitive neuroscience research has witnessed a rapidly growing interest on brain state classification using fMRI and experimental designs.

In 2007, LaConte et al. \cite{43} performed online classification of the left and right index finger movement using SVM, which verified the feasibility of using machine learning to implement a BCI system. They first trained a SVM classifier on offline fMRI data, then applied it to online fMRI images to predict the brain state, and next updated the computer-presented stimulus accordingly. This study also showed that machine learning based stimulus feedback can respond to changes in the brain state much earlier than the time-to-peak limitation of the BOLD response, i.e., the former has higher sensitivity. In 2009 Eklund et al. \cite{44} used a neural network to classify three activities (left hand movement, right hand movement, and resting) from rtfMRI, and then controlled the balance of a virtual reality inverted pendulum. In 2011, Hollmann et al. \cite{45} used relevance vector machine to predict a person's decision in the game. In 2013, Andersson et al. \cite{46} used SVM to classify visuospatial attentions based on the fMRI data collected by an ultrahigh field MRI scanner (7 Tesla). Four subjects succeeded in navigating a robot with virtually no training. Compared with methods based on the local activation of ROIs, MVPA has significantly higher information transfer rate.

\section{Future Developments and Ethical Considerations}

In BCIs, EEG has excellent temporal resolution but poor spatial resolution, whereas fMRI has high spatial resolution and low temporal resolution. Recent advances in sensing hardware have enabled the simultaneous acquisition of EEG and fMRI signals, but sophisticated signal processing and machine learning approaches are still needed to optimally integrate these two modalities to achieve both high temporal resolution and high spatial resolution \cite{47,48,49}. Then, brain stimulation techniques like the transcranial magnetic stimulation (TMS) can be better used to treat brain disorders.

The rapid development of BCIs also raises ethical concerns. Both structural and functional brain signals are related to mental states and traits, which could potentially be used to reveal sensitive private information \cite{2}. So, ethics and regulations are also very important to the healthy development of BCIs.

\section{Conclusions}

This paper has introduced the architecture of rtfMRI based BCI, which includes image acquisition, image preprocessing, image analysis, and feedback. Among them, image preprocessing and analysis are the most important components. Though there have been lots of algorithms for offline fMRI data processing and analysis, how to modify and optimize them for online real-time tasks still calls for more research.

We also reviewed the applications of rtfMRI in BCI, which can be divided into two directions: neuralfeedback and brain state decoding. Both can be of great significance to clinical rehabilitation and cognitive neuroscience research.

\end{document}